\documentstyle[12pt]{article}
\begin{document}

\noindent
{\Large\bf  Acceleration time scale}

\vspace{1mm}
\noindent
{\Large\bf at ultrarelativistic shock waves}

\vspace{5mm}
\noindent
{\large J. Bednarz}
\vspace{5mm}

\noindent
{\it Obserwatorium Astronomiczne, Uniwersytet Jagiello\'{n}ski, }

\noindent
{\it ul. Orla 171, 30-244 Krak\'{o}w, Poland }

\date{}
\vspace{10mm}

\noindent
{\bf Abstract }

\noindent
The first-order cosmic ray acceleration at ultrarelativistic shocks is
investigated using the Monte Carlo method. We apply a method of discrete
particle momentum scattering as a model of particle pitch angle diffusion
to reproduce highly anisotropic conditions at the shock wave. Shocks with
Lorentz factors $\gamma$ up to 320 and varying magnetic field inclinations
$\psi$ are considered. Values of diffusion coefficients upstream
in the point where energy spectral indices stabilize to the limit 2.2 were
calculated. The obtained acceleration time does not depend on shock conditions.

\vspace{5mm}
\noindent
{\bf Introduction. }

\noindent
Till quite recently the discussion of first order Fermi acceleration at
relativistic
shocks was restricted to small Lorentz $\gamma$-factors (for review see
Ostrowski[1] and Kirk[2]). Bednarz and Ostrowski[3] found a simple mechanism
that allows particles to be accelerated at shock waves with large Lorentz
factors. It implies that an inclination of formed spectrum does not depend on
downstream but only on upstream conditions and a shock velocity.
The mechanism is a natural proccess accelerating particles in astronomical
objects where such shocks occur.

Among these phenomena where ultrarelativistic shocks are anticipated are
gamma-ray bursts. They are non-thermal bursts of
low energy $\gamma$-rays. Typical GRB lasts for about 10 sec. Their broad
spectra usually peak between a few 100 keV and a few MeV (for review
see Piran [4]). The Lorentz factor of an expanding fireball larger then 300
is required in order to the pair production optical depth to be
sufficiently small (Baring [5]). Recently the host galaxy with redshift
3.42 was indentified for GRB971214 (Kulkarni et al. [6]) which confirms the
suggestion that they originate at cosmological sources (Meegan et al. [7]).

\vspace{5mm}
\noindent
{\bf Numerical simulations of the acceleration process. }

\noindent
In simulations we follow the procedure used in Bednarz and
Ostrowski [8]. Seed particles are injected downstream the shock
with the same initial energy. Theirs trajectories are derived in
homogenous magnetic field $B_0$ perturbed by inhomogenities which
are simulated by particle momentum scattering within a cone with
small angular opening $\Delta \Omega$ less than the particle anisotropy
$\sim 1/\gamma$ (see Ostrowski [9]). A particle is excluded
from the simulation either if it escapes through the free-escape
boundary placed far downstream the shock or if it reaches a time
or energy larger than the assumed upper limit. These particles are
replaced with ones arising from splitting the remaining high-weight
particles with preserving their physical parameters. They are not
replaced if they reach upper limit of time.

\begin{figure}
\vspace*{5.5cm}
\includegraphics{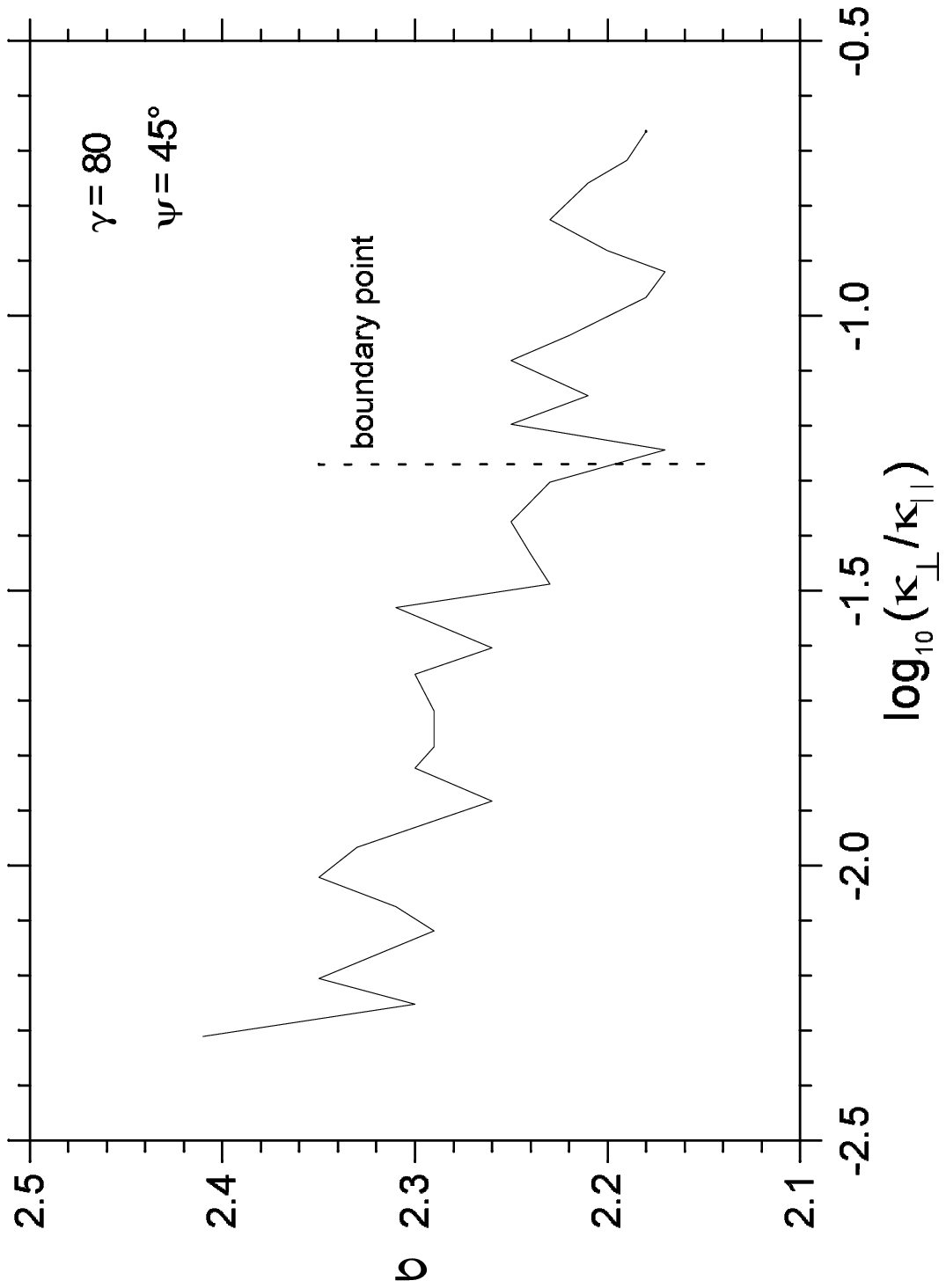}
\caption{
The simulated spectral indices $\sigma$ for particles accelerated at   
shocks with different upstream diffusion represented by
${\rm log}_{10} (\kappa_\perp / \kappa_\| )$. Downstream we assume
$\kappa_\perp = 0$. Perpendicular dashed line points to `boundary point'.}
\label{fig1}
\end{figure}

\begin{figure}
\vspace*{5.5cm}
\includegraphics{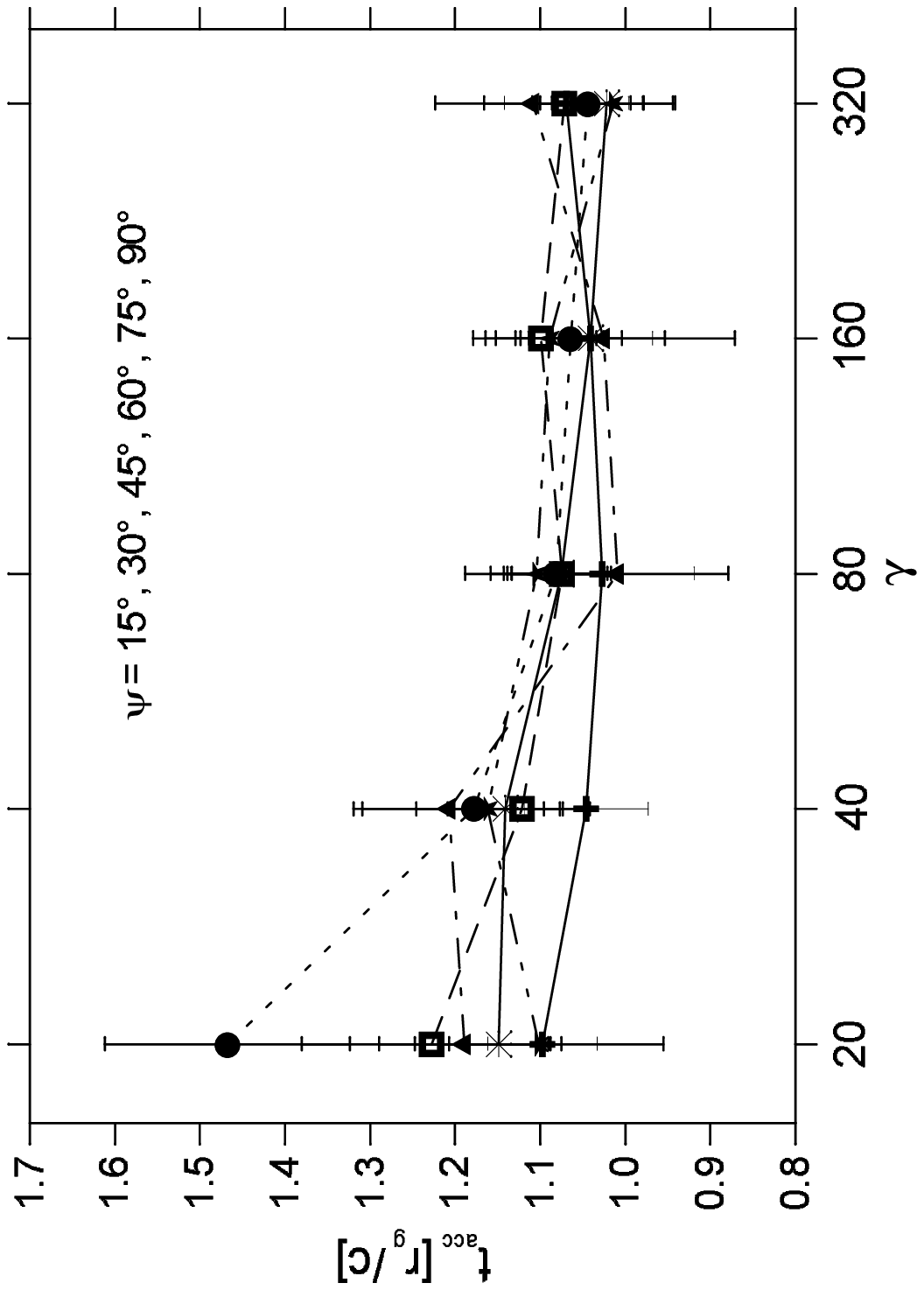}
\caption{
The simulated acceleration times $t_{acc}$ for particles accelerated at   
shocks with different Lorentz factors $\gamma$. The considered upstream
magnetic field inclination are represent by different lines with indicated
simulation errors. The $t_{acc}$ value near 1.5 is for $\psi = 30^\circ$
and results from large simulation error at $\gamma = 20$.}
\label{fig2}
\end{figure}
\noindent
Computations are
performed in the respective - upstream or downstream - plasma rest
frame. If particles cross the shock their parameters are transformed to
the shock rest frame. The respective contribution divided by their
velocity component normal to the shock added to momentum bin
forms particle spectra with shifting cut-off point toward higher
energies. These spectra allow us to derive their inclination and
acceleration time (see Bednarz and Ostrowski [8]). Time measured
in the shock rest frame in downstream $r_{g}/c$ units was transformed
to the downstream plasma rest frame ($r_{g}$ is a particle gyroradius).
That can be done because we know the ratio of mean time that particle
spends upstream to mean time it spends downstream. This value is small
and due to this the time measured in the shock rest frame is of
1.05 factor shorter than downstream for large $\gamma$.

In this paper we simulate different conditions upstream and downstream
the shock, represented by cross-field diffusion coefficient
$\kappa_\perp$ and the parallel diffusion coefficient $\kappa_\|$.

\vspace{5mm}
\noindent
{\bf Results.}

\noindent
We take into consideration shocks with the Lorentz factors equal
$\gamma=$20, 40, 80, 160, 320 and magnetic field inclinations
upstream the shock $\psi=15^\circ, 30^\circ, 45^\circ$, 
$60^\circ, 75^\circ, 90^\circ$.

At first we consider downstream conditions without magnetic
fluctuations. We search by simple data inspection as in
Fig. 1 for a value $\kappa_\perp /\kappa_\|$ where the spectral
index reaches limiting value of 2.2. It yields a fit:

$\kappa_\perp /\kappa_\| = 0.25\cdot \gamma^{-1.22} \psi$

\noindent
The fit is good in full considered parameter space.

The acceleration time $t_{acc}$ for the boundary points is presented
in Fig. 2. where one can seen the lack of any change with $\psi$.
The figure exibits also a slow $t_{acc}$ decrease from 1.2 to approach
the value 1.1 at $\gamma =80$. We observe small tendency of
$t_{acc}$ to grow if $\sigma$ increases to 2.3-2.4 and no further
change if $\kappa_\perp /\kappa_\|$ grows.
In the next step we introduce magnetic field fluctuations
downstream to be $\kappa_\perp /\kappa_\|=7.1\cdot10^{-4}$ or
$1.1\cdot 10^{-1}$. $t_{acc}$ behaves nearly as in the case without
fluctuations, with only a small reduction (circa 0.1) for the larger
fluctuation.
Thus the resulting acceleration times are always approximately
$t_{acc}$ = 1 (in units of $r_{g}/c$ downstream the shock)

\vspace{5mm}
\noindent
{\bf Discussion. }

\noindent
We have obtained the result that is not suprising if we recall
the way particles are accelerated (Bednarz and Ostrowski [3])
and that they do not spend much time upstream.

We expect our results can be applied in modelling relativistic shock
waves expected to occur in gamma-ray burst sources and at pulsar wind
terminal shocks. The model of Kenel and Coroniti [10] assumes for
Crab Nebula initial flow with Lorentz factor equal to $10^{6}$.
One should note that the acceleration time is short, close to the
value of $r_{g}/c$ downstream the shock. However, it is not very
short with energy gains $\sim \gamma^{2}$ at single particle-shock
interaction. It results from the fact that it is not possible to
reflect a particle from the ultrarelativistic shock ( cf., Begelman
and Kirk [11], Ostrowski [1])

\vspace{5mm}
\noindent
{\bf Acknowledgements. }

\noindent
The author is grateful to Micha{\l} Ostrowski for valuable discussions.

\noindent
The present work was supported by the {\it Komitet Bada\'n Naukowych}
through the grant PB 179/P03/96/11.

\noindent
The presented computations were done on the CONVEX Exemplars 
in ACK `CYFRONET' in Krak\'ow.

\end{document}